\documentclass[aps,letterpaper,11pt,nofootinbib,preprint,showpacs,preprintnumbers,amsmath,amssymb]{revtex4-1}%
\usepackage{latexsym}%
\usepackage{epsf}%
\usepackage[hypertex]{hyperref}%
\usepackage{graphicx}
\usepackage{epsfig}%
\usepackage{graphicx}%
\usepackage[lofdepth,lotdepth]{subfig}%
\usepackage{color}

\newcommand{\beq}{\begin{equation}}
\newcommand{\beqn}{\begin{equation}\nonumber}
\newcommand{\eeq}{\end{equation}}
\newcommand{\bea}{\begin{eqnarray}}
\newcommand{\bean}{\begin{eqnarray}\nonumber}
\newcommand{\eea}{\end{eqnarray}}

\begin{document}
\begin{center}
{\bf{\Large Canonical Partition function of Loop Black Holes}}
\bigskip\bigskip

{{Kinjalk Lochan$^{a,}$\footnote{e-mail address: \tt{kinjalk@tifr.res.in}}, Cenalo Vaz$^{b,}$\footnote{e-mail address: \tt{cenalo.vaz@uc.edu}},
}}
\bigskip\bigskip

{\it$^a$Tata Institute of Fundamental Research,\\ Homi Bhabha Road, Mumbai-400005, India\\  
\bigskip
$^b$Department of Physics,}
{\it University of Cincinnati,}\\
{\it Cincinnati, Ohio 45221-0011, USA}
\end{center}
\bigskip\bigskip\bigskip\bigskip
\medskip

\centerline{ABSTRACT}
\bigskip\bigskip

 We compute the canonical partition function for quantum black holes in the approach of Loop Quantum Gravity 
(LQG). We argue that any quantum theory of gravity in which the horizon area is built of non-interacting constituents 
cannot yield qualitative corrections to the Bekenstein-Hawking (B-H) area law, but corrections to the area law can
arise as a consequence additional constraints inducing interactions between the constituents. In LQG this is 
implemented by requiring spherical horizons. The canonical approach for LQG seemingly
favours a logarithmic correction to the B-H law with a coefficient of $-\frac 12$. Our initial calculation of the 
partition function uses certain approximations that, we show, do not qualitatively affect the expression for the 
black hole entropy. We later discuss the quantitative corrections to these results when the simplifying approximations 
are relaxed and the full LQG spectrum is dealt with. We show how these corrections can be recovered to all orders 
in perturbation theory. However, the convergence properties of the perturbative series remains unknown for now.
 \vskip 0.5in
\noindent{\bf Keywords}: Loop Quantum Gravity, Black Hole Thermodynamics, Barbero-Immirizi Parameter.
\vfill\eject

\section{Introduction}

Classical and semi-classical properties of black holes have been studied quite vigorously in past several decades 
\cite{CLSBH}. As a result, we have discovered that some nice thermodynamic properties can be associated with them. 
For example, black holes are now understood to be endowed with a temperature and to possess an entropy characterized by 
their mass and by any other charges that may be associated with them. These properties, although derived on the 
semi-classical level, are expected to be robust enough that explaining them in terms of an ensemble of microstates 
has come to be considered of prime importance for any quantum theory of gravity. Many approaches to quantum gravity, 
like String theory \cite{ST}, Loop Quantum Gravity (LQG) \cite{LQG} and the theory of Causal Sets \cite{CST}, have 
therefore taken up this problem and each has succeeded in some measure in providing some deeper insight into the problem. 
However, the Bekenstein-Hawking area law seems generic enough that all the approaches, although differing considerably in 
design, are able to reproduce it. As a result, there  remains no general consensus on a final theory of quantum gravity  
and it becomes essential to attempt not only to obtain the original semi-classical results, but also to discover, if 
possible, signature corrections of quantum origin to the semi-classical wisdom. 

In this paper, we take up one of the popular approaches to quantum gravity, namely LQG. Black holes have been studied 
quite extensively in this approach \cite{LQGBH1}, mostly in context of isolated horizon scheme \cite{BHasIH}. In 
LQG, the area spectrum of an isolated horizon is readily available and black holes with a given horizon area have 
been studied in the microcanonical ensemble \cite{Rovelli,Kaul,gm05,gm06}. Logarithmic corrections to the 
Bekenstein-Hawking entropy have been shown to exist on the semi-classical level \cite{sclog} and as a signature of LQG \cite{gm05,gm06}. 
They have been proposed in other approaches as well \cite{logterms} but there is some 
confusion about the precise value of the coefficient of this term. Approaches that employ conformal 
symmetry techniques generally seem to prefer the value $-3/2$ whereas the pure quantum geometry approach 
seems to favor the value of $-1/2$.

We analyze LQG black holes in the canonical area ensemble with a finite number of punctures, $N$. The use of 
the area ensemble was first advocated in \cite{Krasnov} and has since been employed by others (see, for eg.
\cite{bv11,bv11a} and references therein). A justification for its use has recently been proposed 
in \cite{fgp11}. The use of $N$ as a statistical variable in the area canonical ensemble was first 
advocated in \cite{m01} and its importance for black holes was emphasized in \cite{gp11}. 
There are similarities between our approach and that of \cite{gm06} and our canonical calculation 
agrees with the results therein, although the latter work was performed in the microcanonical ensemble. The microcanonical and canonical entropies do not have to coincide, of course,
except in the thermodynamic limit. Now it is well known that the microcanonical entropy in LQG 
suffers from the fact that the black hole entropy is not a differentiable function of its arguments but 
rather a ladder (or staircase) function \cite{cdf07,bv11,pp11}. As pointed out in \cite{bv11a}, this 
makes it difficult to interpret basic thermodynamic variables, such as the temperature, which are defined 
in terms of derivatives of the entropy. The canonical partition function is assumed to be, by contrast, 
a smooth function of its arguments. We obtain the canonical partition function without making any 
assumptions concerning our variables or referring to any semi-classical feature, so the following analysis 
is independent of any {\it \`a priori} thermodynamic input. We expect that our approach will provide a 
useful framework for examining area fluctuations about the microcanonical value and for a generalization to 
the grand canonical ensemble, where we should be able to precisely test some of the assumptions of
\cite{gm06,gp11}. Here, we find general agreement with \cite{gm06}. 

In section II we briefly review the LQG approach to black holes. We begin our computation of the canonical partition 
function in section III by making a certain approximation, which we will refer to as the ``shell'' approximation
and define later, to the LQG area spectrum. Ignoring the projection constraint, we easily recover the Bekenstein-Hawking 
area-entropy relation together with an estimate of the Barbero-Immirizi parameter and we show that any correction to 
the Bekenstein-Hawking entropy is inherently absent. In section IV we impose the sphericity (or projection) constraint 
on the black hole horizon and re-evaluate the partition function. The effect of the projection constraint is to produce 
a logarithmic correction to the standard semi-classical result. Section V is devoted to a discussion on relaxing 
the ``shell'' approximation and incorporating the full LQG spectrum in a perturbative fashion. We develop a 
systematic approach to compute the corrections to all orders. We conclude by discussing some issues raised as well 
as the general outlook related to our approach in section VI.
 
\section{Black Holes in LQG}

\begin{figure}
\begin{center}
\includegraphics[scale=0.6]{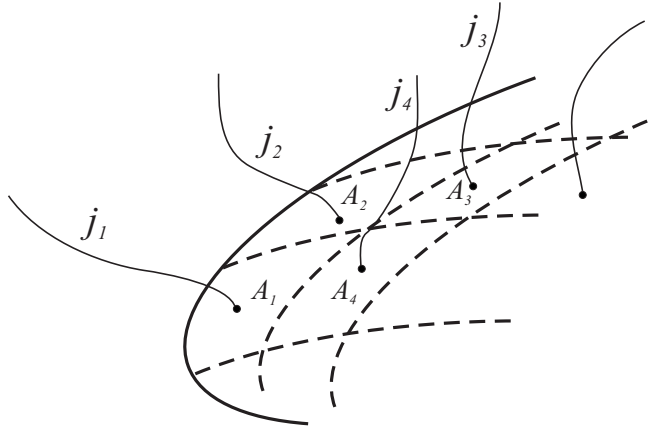}
\caption{Horizon punctured by the edges of a spin-network}
\label{fig-n}
\end{center}
\end{figure}

As a non-perturbative, background independent approach to canonical quantum gravity, Loop Quantum Gravity (LQG) has 
met many challenges effectively, although a full dynamical picture of quantum gravity has remained elusive so far. 
Starting with holonomies, built from {\cal su(2)} valued connections, and fluxes, built from densitized triads, as 
basic variables of the theory, one obtains a well-defined Hilbert space made up of cylindrical functions acting on 
spin-networks over which the holonomies and densitized triads are defined. A spin-network is a graph with 
edges labeled with {\cal su(2)} representations and nodes characterized by  {\cal su(2)} intertwiners. Spin-networks 
are eigenstates of the Area and Volume operators and this leads to a reasonably good description of some of the 
geometrical observables of the theory, at least on the kinematical level. For example, an edge with 
spin-representation $j$ carries an area of eigenvalue 
\begin{equation}
 A_j=8\pi\gamma\l_p^2\sqrt{j(j+1)},
\end{equation}
where $j\in \{\frac 12,1,\frac 32,\ldots\}$ and $\gamma$ is an unknown parameter of quantization known as 
Barbero-Immirizi parameter. Alternatively, an area element of area $A_j$ can be thought as being punctured by an 
edge of the spin-network carrying representation $j$. In general a surface that is punctured by many edges 
of different representations will have the area spectrum 
\begin{equation}
A=8\pi\gamma\l_p^2\sum_{i}\sqrt{j_i(j_i+1)},
\label{areaspectrum}
\end{equation}
where the sum is over all intersections of the edges with the surface. One considers the black hole horizon as an 
isolated horizon \cite{Ashtekar} which is threaded by many edges of a spin-network. Each edge carries some 
representation and hence deposits some area to the horizon as shown in figure \ref{fig-n}. 

In the microcanonical ensemble, one simply tries to count the number of different configurations,
{\it i.e.,} the different ways of threading a horizon with a fixed area $A$ by spin-representations. This then
gives the black hole entropy. However, counting the total number of microstates in this way has not been trivial 
owing to the fact that the area spectrum in \eqref{areaspectrum} is non-distributive in nature. This 
issue has  been discussed and treated in different ways by various authors \cite{Rovelli,gm05,Kaul}. 
In this way, one obtains the asymptotic formula 
\begin{equation}
S_{BH}=\frac{\lambda A(\gamma)}{4l_P^2}, 
\label{BHEnt}
\end{equation}
where we have put the Barbero-Immirzi parameter explicitly in the area as it appears in the expression
\eqref{areaspectrum} for the area, and utilizes the flexibility of fixing $\gamma$ to recover the 
Bekenstein-Hawking result. 

In the following two sections we will examine LQG black holes in the canonical ensemble. We begin by making
the ``shell'' approximation at first, but later, in section V, we generalize to the full LQG spectrum. This 
leads to corrections to the entropy derived in sections III and IV but these corrections amount to a 
renormalization of the Barbero-Immirzi parameter and do not change the functional dependence of the canonical 
partition function on the extensive variables.

\section{Setup for the canonical calculation}

Let us begin by first ignoring the projection constraint and, since all punctures are considered to be distinguishable, 
let us assume that there are $n_j$ punctures carrying spin $j$, then
\beq
A = 8\pi \gamma l_p^ 2 \sum_j n_j \sqrt{j(j+1)} \equiv 8\pi \gamma \sum_j n_j a_j,
\eeq
with $a_j$ capturing the information about the area spectrum. For a configuration containing $N$ 
distinguishable punctures, $n_j$ of which carry spin $j$, we will have
\beq
\frac{N!}{\prod_j n_j!} \prod_j (2j+1)^{n_j}             \nonumber
\eeq
distinct configurations. The degeneracy factor $(2j+1)^{n_j}$ signifies $2j+1$ values for $m_j=\{-j,-j+1,..,j\}$ 
for a spin-representation $j$. Therefore, we may write the partition function as
\beq
Z(\beta,N) = \sum_{\{n_j\}} \frac{N!}{\prod_j n_j!} \prod_j (2j+1)^{n_j} e^ {-8\pi \gamma \beta 
n_j a_j},
\eeq
where $\beta$ is conjugate to the area and is not the black hole temperature. By the Binomial theorem, 
this is
\beq
Z(\beta,N) = \left(\sum_j (2j+1) e^ {-8\pi\gamma \beta  a_j} \right)^N.
\eeq
Now, since $j \in \{\frac 12,1,\frac 32, \ldots \}$ it is convenient to make the change $2j = l$ 
so that 
\beq
Z(\beta,N) =  \left(\sum_{l=1,2,\ldots}(l+1)e^{-\sigma \sqrt{l(l+2)}}\right)^N = z^N(\sigma)
\label{withnosph}
\eeq
where $\sigma = 4\pi \gamma l_p^2\beta $. We will now study the general properties of this kind of 
partition function without restricting ourselves to to this or any particular area spectrum.

\subsection{Formal Solution}

\noindent If, as we have above,
\beq
Z(\sigma,N) = z^N(\sigma)
\label{firstform}
\eeq
then, because $A = -\frac{\partial\ln Z}{\partial\beta}$, it follows that
\beq
\frac A{4\pi\gamma l_p^2 N} = -\frac{\partial \ln z(\sigma)}{\partial \sigma}~ 
\stackrel{\text{def}}{=}~ q. \label{qDEF}
\eeq
Inverting this relation, one finds a solution of the form
\beq
\sigma = 4\pi \gamma l_p^2 \beta = \sigma(q).
\eeq
The entropy of the system is obtained from
\beq
S = \ln Z(\sigma,N) + \beta A(\sigma,N) = N [\ln z(\sigma) + \beta a(\sigma)]
\eeq
where, $\beta = \beta(q)$ as above and $A=Na$ {\it i.e.,} $a = 4\pi\gamma l_p^2 q,$ as can 
be seen from (\ref{qDEF}). Thus,
\beq
S(q) = N \left[\ln z\{\sigma(q)\} + q\sigma(q)\right]. 
\label{Entropy}
\eeq
Now let us maximize the entropy with respect to the total number of constituents (punctures
for the LQG model), $N$. Then
\beq
\frac SN + N\left(\frac{\partial \ln z}{\partial \sigma} \sigma' + \sigma + q \sigma'\right)
\frac{\partial q}{\partial N} = 0,
\eeq
where the prime refers to a derivative with respect to $q$. Now from (\ref{qDEF}),
\beq
\frac{\partial q}{\partial N} = - \frac A{4\pi\gamma l_p^2 N^2} = - \frac qN,
\eeq
so,
\beq
\frac SN - q\left[\frac{\partial \ln z}{\partial \sigma} \sigma' + \sigma + q\sigma'\right]=0,
\eeq
and, since
\beq
\frac{\partial \ln z}{\partial \sigma} = - \frac A{4\pi \gamma l_p^2 N} = - q,
\eeq
we have
\beq
\frac SN - q\sigma(q) = 0
\eeq
or
\beq
\ln z\{\sigma(q)\} = 0 \Rightarrow z\{\sigma(q)\} = 1.
\eeq
We must solve this equation for $N=N(A)$ and reinsert into $S$ to determine the entropy as a function 
of area. In this way we will get
\beq
S = \left. N(A) q \sigma(q) \right|_{z\{\sigma(q)\} = 1},
\eeq
but the relation $z\{\sigma(q)\} = 1$ implies that $q=q_0$, a constant. This gives
\beq
N(A) = \frac A{4\pi \gamma l_p^2 q_0},
\eeq
and
\beq
S = \sigma(q_0) \frac A{4\pi \gamma l_p^2}.
\label{entshells}
\eeq
The entropy is always proportional to the area so long as the canonical partition function obeys \eqref{firstform} 
and regardless of the spectrum. It follows that any theory of quantum gravity in which non-interacting constituents 
make up the horizon (like the punctures in LQG at this level) will result in the Bekenstein-Hawking area-entropy 
law. To obtain the Bekenstein-Hawking relation in LQG framework, the value of the Immirizi parameter should be 
fixed at
\beq
\gamma = \frac{\sigma(q_0)}\pi. \label{BI}
\eeq
Within this framework, any correction (in particular logarithmic) must arise from additional 
constraints (such as the projection constraint in LQG) on the horizon. Therefore, we do not expect 
any logarithmic correction to Bekenstein-Hawking relation in shell collapse scenarios \cite{vw01}. 
Now we estimate the numerical value for $\gamma$ for  an approximated LQG spectrum, which 
looks very much like the spectrum in dust shell collapse. 

\subsection{The `Shell'' approximation to the LQG spectrum}

We define the ``shell'' approximation by $\sqrt{l^2+2l} = \sqrt{(l+1)^2-1} \approx l+1$. We have called this
the ``shell'' approximation because an identical solution is obtained in the canonical quantization of the 
LTB dust models \cite{vw01}. In those models the collapsing dust ball is viewed as made up of dust shells,
and the apparent horizon for each shell is shown to have a similar area spectrum. In making this approximation, 
we will be making errors towards low spin punctures, but for large spin punctures, the errors will not be that 
significant. Then, 
\beq
z(\sigma) = \sum_{l=1,2,\ldots} (l+1) e^{-\sigma(l+1)} = -\frac{\partial}{\partial\sigma}
\sum_{l=0}^\infty e^{-\sigma(l+2)} = \frac{2-e^{-\sigma}}{(e^\sigma-1)^2},
\label{zs}
\eeq
so that the condition $z\{\sigma(q)\} = 1$ translates to
\beq
2- e^{-\sigma} = (e^\sigma-1)^2.
\eeq
There are two possible solutions for $e^\sigma$, {\it viz.,}
\beq
e^\sigma = \left\{\begin{matrix}
2.247\cr
0.555\end{matrix}\right.
\eeq
and the equation for $q$
\beq
q = - \frac{\partial \ln z}{\partial\sigma} = 2+ \frac 2{e^\sigma-1} + \frac 1{1-2e^\sigma},
\eeq
determines $q_0$ in terms of these solutions. There is only one positive solution for $q$ and it corresponds to 
$e^\sigma = 2.247$; we end up with
\beq
\sigma(q_0) = 0.810,~~ q_0 = 3.318,~~ \gamma = 0.258,
\eeq
using (\ref{BI}).
It is important to note that even if an exact calculation could be done then the entropy would still be proportional 
to $A$ while the constants $q_0$ and $\gamma$ would change. Now, for the estimate of the Barbero-Immirizi parameter 
from microcanonical counting, Ghosh and Mitra, in their study of LQG black holes \cite{gm05}, suggest that 
the correct value for $\gamma$ is $0.274$. On the other hand, Ling and Zhang produce a very similar value, 
$\gamma \approx 0.247$, from their study \cite{LZ} of $N=1$ super-symmetric LQG black holes. Importantly, this value 
of the parameter $\gamma$ is closely related to the agreement between LQG spectrum and quasi-normal (ringing-) 
mode frequency of black holes \cite{Rovelli}. One can in principle, get a better estimate for $\gamma$ by dropping 
the ``shell'' approximation. We will illustrate how this can be done in section V. 

For the present we recall that the projection constraint is still missing and in the next section we will introduce 
it. This constraint has been identified \cite{gm06} as the possible source of a logarithmic correction 
from  analyses in the microcanonical ensemble. First we write the canonical partition function with this constraint, but 
still staying within the ``shell'' approximation.

\section{The Projection Constraint}

We now consider a spherically symmetric black hole. Such a system in LQG will be described by a 
spin-network which is an eigenstate of the projection operator $\hat{n}\cdot\vec{J}$ with zero eigenvalue.
In other words, the $m_j$ values carried by the spins puncturing the horizon must add up to zero. 
This is the projection constraint. It can be written as,
\beq
\sum_{j, m_j}n_{jm_j} m_j = 0,
\eeq
where $n_{jm_j}$ gives the number of punctures carrying spin $j$ and projection $m_j = -j,-j+1,\ldots,
j-1,j$ along the $z-$axis. We can see that the punctures are now ``interacting'' in the sense they have to satisfy
a constraint for getting the desired quantum state and their 
interaction will manifest itself in terms of some correction to \eqref{entshells}.

First we write,
\beq
2\sum_{j, m_j}n_{jm_j} m_j = p,
\eeq
for some integer $p,$ then we will put the constraint on $p$, forcing it to vanish.
Suppose there are $N=\sum_{j,m_j} n_{jm_j}$ 
punctures, then the partition function becomes
\beq
Z(\beta,N) = \sum_{\{n_{jm_j}\}} \frac{N!}{\prod_{jm_j} n_{jm_j}!}~\delta_{p,0}~ e^{-8\pi\gamma\beta \sum_{jm_j} n_{jm_j}a_j},
\eeq
where $\delta_{p,0}$ is the Kronecker delta function. We want a suitable representation for this
function. For integer $p$ a convenient representation is\footnote{To see this, note that the right hand side is 
\beqn
\left.\frac 1{2\pi} \frac{e^{ikp}}{ip}\right|_0^{2\pi} = \frac 1{2\pi i p} \left[e^{2\pi i p} - 1\right].
\eeq
Since $p\in \mathbb{Z}$, if $p\neq 0$ then the right hand side is vanishing. In the limit as $p\rightarrow 0$,
this becomes
\beqn
\lim_{p\rightarrow 0} \frac{\sin(2\pi p)}{2\pi p} = 1,
\eeq
so the representation is true.}
\beq
\delta_{p,0} = \frac 1{2\pi} \int_0^{2\pi} dk~ e^{ikp}.
\eeq
and using it we find
\beq
Z(\beta,N) = \frac 1{2\pi} \sum_{\{n_{jm_j}\}} \frac{N!}{\prod_{jm_j} n_{jm_j}!}\left(\int_0^{2\pi} 
dk e^{2ik\sum_{jm_j} n_{jm_j} m_j}\right) e^{-8\pi\gamma\beta \sum_{jm_j} n_{jm_j}a_j}.
\eeq
Interchanging the sum over $n_{jm_j}$ and the integral over $k$,
\bea
Z(\beta,N) &=& \frac 1{2\pi} \int_0^{2\pi} dk \sum_{\{n_{jm_j}\}} \frac{N!}{\prod_{jm_j} n_{jm_j}!} 
e^{2ik\sum_{jm_j}n_{jm_j} m_j} e^{-8\pi\gamma\beta \sum_{jm_j} n_{jm_j}a_j},\cr\cr
&=& \frac 1{2\pi} \int_0^{2\pi} dk \left(\sum_{jm_j}e^{(2ik m_j - 8\pi\gamma\beta a_j)}\right)^N,
\eea
again after using the Binomial theorem. Note that to recover the result with no projection constraint, 
we must drop the integration over $k$ and set $k=0$. Now let $2j=l$ as before, $2m_j = r_l \in \{-l,-l+2,\ldots,
l-2,l\}$ and consider the sum in the integrand first,
\beq
\sum_{l=1}^\infty \left(\sum_{r_l}e^{ik r_l}\right)e^{-\sigma\sqrt{l(l+2)}} = \sum_{l=1}^\infty 
\frac{e^{ik(l+2)}-e^{-ikl}}{e^{2ik}-1} e^{-\sigma\sqrt{l(l+2)}}.
\label{withsph}
\eeq
We easily check that the limit as $k\rightarrow 0$ of the sum is 
\beq
\lim_{k\rightarrow 0} \frac{e^{ik(l+2)}-e^{-ikl}}{e^{2ik}-1} = l+1,
\eeq
and therefore \eqref{withsph} reduces to \eqref{withnosph}. 

We want to get the partition function
\bea
Z(\beta,N) &=& \frac 1{2\pi} \int_0^{2\pi} dk \left(\frac 1{e^{2ik}-1}\sum_{l=1}^\infty e^{-\sigma\sqrt{l(l+2)}}
\{e^{ik(l+2)} -e^{-ikl}\}\right)^N,\cr\cr
&\approx& \frac 1{2\pi} \int_0^{2\pi} dk\left(\frac 1{e^{2ik}-1}\sum_{l=1}^\infty e^{-\sigma(l+1)}\{e^{ik(l+2)}
-e^{-ikl}\}\right)^N,\cr\cr
&=& \frac 1{2\pi} \int_0^{2\pi} dk \left(\frac 1{e^{2ik}-1}\sum_{l=0}^\infty e^{-\sigma(l+2)}\{e^{ik(l+3)}
-e^{-ik(l+1)}\}\right)^N,
\eea
into a manageable form. In the second step, we made use of shell approximation to simplify the 
calculation. After a little manipulation, we end up with the expression 
\beq
Z(\beta,N) =  \frac 1{2\pi} \int_{-\pi}^{\pi} dk \left(\frac{2\cos k-e^{-\sigma}}{e^{2\sigma}-2e^\sigma \cos 
k + 1} \right)^N,
\eeq
where we have made use of the evenness of the integrand to rewrite the limits of the integration. Note that this 
is not of the form \eqref{firstform} and that this is because of the projection constraint, which introduced an 
interaction between punctures as discussed above. In the limit as $k\rightarrow 0$, the integrand 
approaches
\beq
\frac{2-e^{-\sigma}}{(e^\sigma-1)^2},
\eeq
which is precisely \eqref{zs} as required. 

Now we try to approximate the integral from the information we have about the integrand. First, we see 
that it is a unimodal symmetric distribution between its limits. So we would like to approximate it by 
a ``bell-curve''. (In order to do so, we can make the transformation
\beq
k=2\tan^{-1}(x/2),
\eeq
to get the limits form $-\infty$ to $\infty$.) We then make an assumption that it {\it is} well-approximated 
by a Gaussian in some regime of $N$. If true, then the variance ($\tilde{\sigma}^2$) of the ``bell-curve'' 
is obtained by the second derivative of the integrand at the peak, since in the case of a Gaussian
\beq
\tilde{\sigma}^2=-\frac{f(x)|_0}{f''(x)|_0}.
\eeq
In our case,
\begin{eqnarray}
f(x)= \frac 1{2\pi} \left(\frac{2\cos k(x)-e^{-\sigma}}{e^{2\sigma}-2e^\sigma \cos k(x) + 1}
\right)^N \left|\frac{dk}{dx}\right|, \nonumber\\
=\frac 1{2\pi(1+x^2/4)} \left(\frac{2\cos k(x)-e^{-\sigma}}{e^{2\sigma}-2e^\sigma \cos k(x) + 1}
\right)^N,
\end{eqnarray}
resulting in the variance
\beq
\tilde{\sigma}^2=-\frac{f(x)|_0}{f''(x)|_0}=\frac{2(e^\sigma-1)^2(2e^{\sigma}-1)}{-1+e^\sigma(4+e^\sigma(-5+e^{\sigma}(2+4N)))},
\eeq
which in large $N$ limit tends to 
\beq
\lim_{N\rightarrow \infty}\tilde{\sigma}^2\rightarrow \frac{(e^\sigma-1)^2(2-e^{-\sigma})}{2e^{2\sigma}N}.
\eeq
Now, being a symmetric distribution its skewness is already zero. We analyze its kurtosis next. For 
a Gaussian with a zero mean, the kurtosis obtained from fourth standardized moment, can be given by its distribution function as
\beq
\beta_2=\frac{\mu_4}{\tilde{\sigma}^4}=\frac{f(x)|_0f^{(4)}(x)|_0}{[f''(x)|_0]^2}=3,
\eeq
which gives the ``excess kurtosis'' as
\beq
\beta_2-3=0.
\eeq
For our case, also due to the fact that its mean is zero, we obtain,
\beq
\frac{\mu_4}{\tilde{\sigma}^4}=\frac{f(x)|_0f^{(4)}(x)|_0}{[f''(x)|_0]^2}=\frac{6[(1-2e^\sigma)^2
(e^\sigma-1)^4 + 8e^{3\sigma}(-1 + e^\sigma(2 + e^\sigma (e^\sigma-1)))N+8e^{6\sigma}N^2]}{[-1+
e^\sigma(4+e^\sigma(-5+e^{\sigma}(2+4N)))]^2}
\eeq
and we see that in limit of large $N$, the excess kurtosis of this distribution also vanishes,
\beq
\lim_{N\rightarrow \infty}\frac{\mu_4}{\tilde{\sigma}^4}-3\rightarrow 0. 
\eeq
For large $N$ the integrand tends to a unimodal symmetric distribution with zero skewness and 
vanishing excess kurtosis. Thus, our initial assumption is justified and in the large $N$ 
limit this function is indeed well approximated by a Gaussian. In this limit 
the partition function is given by the area of a gaussian, which can be readily evaluated. 
Therefore, we have the partition function in the thermodynamic limit
\beq
Z(N,\sigma) = f(x)|_0\sqrt{2\pi} \tilde{\sigma}.
\eeq
or, explicitly,
\beq
Z(N,\sigma) = \sqrt{\frac{(e^\sigma-1)^2(2-e^{-\sigma})}{2e^{2\sigma}N}}
\left(\frac{2-e^{-\sigma}}{(e^\sigma-1)^2}\right)^N,
\eeq
and thus,
\beq
\ln Z \approx N \ln z(\sigma) - \frac 12 \ln N + \text{const.},
\eeq
(assuming that $N$ is large) where $z(\sigma)$ is given in \eqref{zs}. We find, as before, 
\beq
\frac A{4\pi\gamma l_p^2 N} =  -\frac{\partial \ln z}{\partial \sigma}~ \stackrel{\text{def}}{=}~ q,
\eeq
and 
\beq
S = \ln Z + \beta A = N [\ln z(\sigma)+\sigma q] - \frac 12 \ln N + \text{const.},
\eeq
Now
\beq
\frac{\partial S}{\partial N} = \ln z - \frac 1{2N} \approx \ln z = 0. 
\eeq
for large $N$ and retaining terms up to the order of $1/\sqrt{N}$. This implies, once again, that $z(\sigma)=1,$ and it 
can be solved for $q_0$ and $\sigma(q_0)$ as before. Then
\beq
S \approx q_0 \sigma(q_0) N(A) - \frac 12 \ln N(A) + \text{const.},
\eeq
with $N = A/4\pi \gamma l_p^2 q_0$, {\it i.e.,}
\beq
S \approx \frac{\sigma(q_0) A}{4\pi\gamma l_p^2} - \frac 12 \ln\left(\frac{A}{4\pi\gamma l_p^2 
q_0}\right) + \text{const.},
\eeq
or with the chosen value of the Immirizi parameter
\beq
S = \frac{A}{4l_p^2} - \frac 12 \ln \left(\frac{A}{4l_p^2}\right) + \text{const.}
\eeq
Thus we obtain a logarithmic correction, whose origin lies clearly in the imposition of the projection 
constraint. The logarithmic correction comes with a factor $-\frac{1}{2}$, as already pointed out in 
\cite{gm05,gm06}, and therefore the canonical calculations are shown agree with the microcanonical results.
The microcanonical ensemble and the canonical ensemble agree even at the subdominant correction, in contrast 
with the results of \cite{bv11a}, in which the coefficient is determined to be $+\frac 12$. The reason 
for this discrepancy is that the authors in \cite{bv11a} work in the grand canonical ensemble (with a 
vanishing chemical potential). Fluctuations in the number of punctures can be shown to contribute 
precisely a logarithmic term with coefficient $+1$ to the entropy. The sum of this contribution 
and the logarithmic term from the projection constraint in the canonical ensemble above leads to their 
result. The projection constraint, absent in the shell-picture, plays a pivotal role in bringing about 
the LQG signature.

\section{Towards a full $z(\sigma)$ }

\noindent We will now try to see how the full LQG spectrum is likely to improve upon the calculational 
details. In the previous sections, we made the approximation
\beq
\sqrt{l(l+1)} = \sqrt{(l+1)^2-1} \approx l+1,
\eeq
in evaluating the partition function. As already argued, this type of approximation will induce a 
significant error only towards low spin punctures. For a more precise calculation, we actually 
want
\beq
z(\sigma) = \sum_{l=1,2,\ldots} (l+1) e^{-\sigma\sqrt{(l+1)^2-1}},
\eeq
which can be re-written as 
\beq
z(\sigma) = \sum_{l=2,\ldots} (l) e^{-\sigma\sqrt{(l)^2-1}},
\eeq
The most direct way to evaluate the sum above is to employ the Mellin-Barnes representation of 
the exponential function,
\beq
e^{-\alpha} = \frac 1{2\pi i} \int_{\tau-i\infty}^{\tau+i\infty} ds~ \Gamma(s)~ \alpha^{-s}
\eeq
where $\tau \in \mathbb{R}^+$, $\text{Re}(\alpha)>0$ and the integral is taken over a line parallel 
to the imaginary axis. The integration path can be closed in the left half plane.

Using this representation we find
\bea
z(\sigma) &&= \sum_{l=2}^\infty \frac 1{2\pi i} \int_{\tau-i\infty}^{\tau+i\infty} ds~ \Gamma(s)~
\sigma^{-s}l^{-s+1}\left[1-\frac 1{l^2}\right]^{-s/2},\cr\cr
= && \sum_{l=2}^\infty \frac 1{2\pi i} \int_{\tau-i\infty}^{\tau+i\infty} ds~ \Gamma(s)~
\sigma^{-s} l^{-s+1} \left[1+\sum_{k=0}^\infty l^{-2(k+1)} \frac{s(s+2)\ldots (s+2k)}{2\times 4\cdot 
\ldots \times 2(k+1)}\right],\cr\cr
= && \sum_{l=2}^\infty \frac 1{2\pi i} \int_{\tau-i\infty}^{\tau+i\infty} ds~ \Gamma(s)~
\sigma^{-s} l^{-s+1}\cr \cr
&&\hskip 1in +\sum_{k=0}^\infty \frac 1{2\pi i} \int_{\tau-i\infty}^{\tau+i\infty} ds~ 
\Gamma(s)~ \sigma^{-s} \{\zeta(s+2k+1)-1\}\left[\frac{s(s+2)\ldots (s+2k)}{2\times 4\cdot 
\ldots \times 2(k+1)}\right].\cr
&&
\eea
In the above equations $\zeta(s)$ is the Riemann zeta-function and $\Gamma(s)$ is the Gamma function.
Now we may systematically expand, order by order in $k$, to estimate the possible corrections.

\subsection{Shells again}

It is easy to verify that the first sum yields the ``shell'' approximation, for it can be written 
in the form
\beq
z(\sigma) \approx\frac 1{2\pi i} \int_{\tau-i\infty}^{\tau+i\infty} ds~ \Gamma(s)~ \sigma^{-s} 
\{\zeta(s-1)-1\},
\eeq
which has simple poles at $s = \{2,0,-1,\ldots\}$, so has the value
\beq
= \frac 1{\sigma^2} + \sum_{n=0}^\infty \frac{(-1)^n \sigma^{n}}{n!}\{\zeta(-n-1)-1\},
\eeq
using residue theorem. Now, using the relation
\beq
\zeta(-n) = -\frac{B_{n+1}}{n+1},
\label{zetabern}
\eeq
between Riemann zeta-function and the Bernoulli numbers, we get
\beq
= \frac 1{\sigma^2} - \sum_{n=0}^\infty \frac{(-1)^n \sigma^{n}}{n!}\left\{\frac{B_{n+2}}{n+2}+1\right\}
\eeq
and again, starting with the expression
\beq
\frac 1{1-e^{-\sigma}} = \sum_{n=0}^\infty \frac {(-1)^n  B_n\sigma^{n-1}}{n!} = \frac{B_0}\sigma - 
B_1 + \sum_{n=2}^\infty \frac {(-1)^n B_n\sigma^{n-1}}{n!},
\label{expr1}
\eeq
we find
\beq
\sum_{n=0}^\infty \frac {(-1)^n B_{n+2}\sigma^n}{(n+2)n!} = \frac 1{\sigma^2} + \frac{\partial}
{\partial \sigma}\left(\frac 1{1-e^{-\sigma}}\right),
\eeq
and therefore
\bea
z(\sigma) &\approx& - \frac{\partial}{\partial \sigma}\left(\frac 1{1-e^{-\sigma}}\right) - e^{-\sigma} = 
\frac{2-e^{-\sigma}}{(e^\sigma-1)^2} + \ldots
\eea
This is precisely what we have been using.

\subsection{First correction}

To the above we must add the infinite series in $k$. For $k=0$ we have two terms
\beq
\frac 1{4\pi i} \int_{\tau-i\infty}^{\tau+i\infty} ds~ s\Gamma(s)~ \sigma^{-s} \zeta(s+1) -
\frac 1{4\pi i} \int_{\tau-i\infty}^{\tau+i\infty} ds~ s\Gamma(s)~ \sigma^{-s}. 
\eeq
The first term has a simple pole at $s=0$ with residue $+1$ and also simple poles at $s=-n \in 
\mathbb{Z}^-$ with residues
\beq
r_n=\frac{(-1)^{n-1}\zeta(1-n)\sigma^n}{(n-1)!},
\eeq
so, it is 
\bea
&=& \frac 12 \left[1+\sum_{n=1}^\infty \frac{(-1)^{n-1}\zeta(1-n)\sigma^n}{(n-1)!}\right] = 
\frac 12 \left[1 + \sum_{n=1}^\infty\frac{(-1)^n B_n\sigma^n}{n!}\right],\cr\cr
&=& \frac \sigma{2(1-e^{-\sigma})},
\eea
using \eqref{zetabern} and \eqref{expr1}. The second of these is also straightforward since the poles 
are only those of $s\Gamma(s)$, which lie at $\{-1,-2,\ldots,-n,\ldots\}$ with residues $(-1)^{n-1}/(n-1)!$ so
\beq
\frac 1{4\pi i} \int_{\tau-i\infty}^{\tau+i\infty} ds~ s~\Gamma(s)~ \sigma^{-s}  = 
\frac 12 \sum_{n=1}^\infty \frac{(-1)^{n-1} \sigma^n}{(n-1)!} = \frac \sigma 2 \sum_{n=0}^\infty 
\frac{(-1)^n \sigma^n}{n!} = \frac \sigma 2 e^{-\sigma}.
\eeq
Combining the two, this correction is 
\beq
\frac \sigma 2\left[\frac{1-e^{-\sigma}+e^{-2\sigma}}{1-e^{-\sigma}}\right],
\eeq
and so
\beq
z(\sigma) \approx \frac{2-e^{-\sigma}}{(e^\sigma-1)^2} + \frac \sigma 2\left[\frac{1-e^{-\sigma}+e^{-2\sigma}}
{1-e^{-\sigma}}\right] +\ldots
\eeq
The condition $z(\sigma)=1$ becomes a transcendental equation and cannot be solved in closed form but one can 
numerically estimate the left hand side and we find only one positive solution for $q_0$,
\beq
\sigma_(q_0) \approx 1.187,~~ q_0 \approx 0.510,~~ \gamma \approx 0.378.
\eeq
This gives $\sigma(q_0)$ and hence $\gamma$ values that are somewhat higher than in the previous order. Obviously, the 
series oscillates about its asymptotic value so an important question to ask is: how fast does the series converge? We 
will deal with convergence in a subsequent work.

\subsection{$k>0$}
Each of the next correction terms would be of the form
\beq
 \frac 1{2\pi i} \int_{\tau-i\infty}^{\tau+i\infty} ds~ 
\Gamma(s)~ \sigma^{-s} \{\zeta(s+2k+1)-1\}\left[\frac{s(s+2)\ldots (s+2k)}{2\times 4\cdot 
\ldots \times 2(k+1)}\right]
\eeq
with $k \in \mathbb{N}$. For example, at order $k=1$, the correction term would be
\beq
\frac 1{16\pi i} \int_{\tau-i\infty}^{\tau+i\infty} ds~ s(s+2)\Gamma(s)~ \sigma^{-s} \zeta(s+3)-
\frac 1{16\pi i} \int_{\tau-i\infty}^{\tau+i\infty} ds~ s(s+2)\Gamma(s)~ \sigma^{-s}, 
\eeq
which can be evaluated as before. The first integral has simple poles at $s = -n \in \{-1,-2,\ldots\}$. with residues 
\beq
\frac{(-1)^n n(n-2)\zeta(3-n)\sigma^n}{n!},~~ n \neq 2,
\eeq
and residue $-\sigma^2$ at $n=2$. The second integral has simple poles at $s = - n \in \{-1,-2,\ldots\}$ with residues
\beq
\frac{(-1)^n n(n-2)\sigma^n}{n!},
\eeq
and so on and so forth. All the higher order corrections can be computed and subsequently added 
following similar calculations. 

It is hoped that the larger the order, the better will be the estimate, but this would have to be 
rigorously shown from a convergence analysis. This issue is not taken in this paper, but will be 
discussed separately elsewhere. 

\section{Conclusion and Discussion}

In this work, we have analyzed quantum black holes in the framework of the canonical ensemble. We 
first obtained the Bekenstein Hawking relation and demonstrated that this relation is a generic 
feature of any quantum gravity theory built around non-interacting constituents of horizon area. 
Such theories will not show any significant departure from the semi-classical analysis, 
apart from any numerical estimates of parameters involved in the quantization scheme. We also studied 
black holes in the Loop Quantization scheme under some simplifying approximations (the ``shell'' 
spectrum) whose effects are expected to be smeared in case of large black holes. For those black 
holes we determined the Barbero-Immirizi parameter to be quite close to one found in literature, 
relating it to the ringing mode of black holes. 

For the particular case of LQG, it was shown unambiguously that the theory predicts a logarithmic
correction to the Bekenstein-hawking entropy and that this correction is a direct consequence of 
spherical horizons. Our computation leads to a logarithmic correction with a factor of $-1/2$ 
and not $-3/2$ as has been suggested in the literature \cite{logterms}. The reason for this is that 
we are working with the reduced $U(1)$ Chern-Simons theory for which there is one projection 
constraint. The factor of $-3/2$ is obtained in models which employ the full $SU(2)$ symmetry 
in which there are three constraints \cite{bkm09}. The factor of $-1/2$ agrees with the results of
\cite{gm05,gm06}. 

The partition function for the full LQG area spectrum was obtained as an infinite series of terms, making 
use of the Mellin-Barnes representation of the exponential function, and corrections at various orders 
were discussed. Convergence of the series remains an open question for now and will be tackled in a future
publication. Interestingly, this approximation scheme was avoided in \cite{abbdv10} with the use of Pell equation. The exact calculation will help understand the convergence of the series and the full and exact
subleading character of the entropy. We are presently examining this possibility and will report on our 
results elsewhere.
\bigskip\bigskip

\noindent{\bf ACKNOWLEDGEMENTS}

\bigskip

\noindent CV is grateful to T.P. Singh and to the Tata Institute of Fundamental Research for their 
hospitality during the time this work was completed. KL and CV acknowledge useful conversations with 
T.P. Singh. We are also grateful for enlightening correspondence with J. F. Barbero G.  This research
was supported in part by the Templeton Foundation under Project ID $\#$ 20768.

\end{document}